\documentclass[letterpaper]{article}

\usepackage[T1]{fontenc}
\usepackage{geometry}
\usepackage{setspace}

\usepackage[style=chem-acs,
			sorting=none,
			articletitle=true]{biblatex}
\usepackage{graphicx}
\usepackage{float}
\newfloat{scheme}{htbp}{los}
\floatname{scheme}{Scheme}
\floatname{chart}{Chart}
\newfloat{graph}{htbp}{loh}

\usepackage{amsmath,amssymb}
\usepackage{chemformula}
\usepackage[version=4]{mhchem}

\usepackage{xcolor}
\usepackage{hyperref}
\hypersetup{
  colorlinks=true,
  linkcolor=blue,
  citecolor=blue,
  urlcolor=blue
}
\usepackage[normalem]{ulem}

\usepackage{xcolor}
\usepackage[normalem]{ulem}

\usepackage{authblk}

\title{Ultrabroadband, high-sensitivity time-resolved terahertz spectroscopy enabled by a Yb-based laser system}

\author[1,2]{Hong-Yuan Wu}
\author[1,2]{Si-Jia Li}
\author[1]{Jing-Long Ma}
\author[1,2,4]{Guo-Qian Liao}
\author[3]{Alexander Holm Ohrt}
\author[3]{Long Cheng}
\author[3]{Peter Uhd Jepsen}
\author[3]{Bin-Bin Zhou}
\author[1]{Si-Jie Zhang*}
\author[1,2,4]{Yu-Tong Li*}

\affil[1]{Beijing National Laboratory for Condensed Matter Physics, Institute of Physics, Chinese Academy of Sciences, Beijing 100190, China}
\affil[2]{School of Physical Sciences, University of Chinese Academy of Sciences, Beijing 100049, China}
\affil[3]{Department of Electrical and Photonics Engineering, Technical University of Denmark, Kongens Lyngby, DK-2800, Denmark}
\affil[4]{Attosecond Science Center, Songshan Lake Materials Laboratory, Dongguan, Guangdong 523808, China}

\date{*Corresponding authors: \href{mailto:sjzh@iphy.ac.cn}{sjzh@iphy.ac.cn}; \href{mailto:ytli@iphy.ac.cn}{ytli@iphy.ac.cn}}

\begin{document}

\maketitle

\begin{abstract}
Ultrafast optical fields have introduced numerous novel light-induced states in quantum materials. Identifying these states and elucidating their formation mechanisms require comprehensive, quantitative access to nonequilibrium electrodynamics across relevant energy scales. Here, we develop a Yb-laser-based time-resolved terahertz spectroscopy (TRTS) platform combining multi-pass-cell pulse compression, spintronic terahertz generation, and air-biased coherent detection. A single time-domain waveform scan provides continuous phase-resolved spectral coverage to 25~THz ($\sim$100~meV) with a Fourier resolution of 0.1~THz ($\sim$0.4~meV). This capability enables the characterization of both low-frequency charge dynamics and higher-energy resonant excitations in quantum materials under a common temporal reference. The system achieves a single-scan time-domain signal-to-noise ratio of 72~dB with long-term waveform stability, while preserving most of the laser pulse energy for tunable excitation. Measurements on intrinsic GaAs resolve extremely weak phonon features above 23~THz and demonstrate quantitative retrieval of equilibrium and transient complex optical properties. The resulting platform can provide a more complete microscopic picture of light-induced states in quantum materials.

\end{abstract}

\section{Introduction}

Ultrafast optical fields have opened new routes to controlling quantum materials, enabling light-induced phase transitions and states beyond those accessible in equilibrium~\cite{RevModPhys.93.041002}. Establishing the nature of light-induced states and the mechanisms underlying their formation requires resolving the evolution of their low-energy electrodynamic response. Time-resolved terahertz spectroscopy (TRTS) provides direct, contact-free access to this information through phase-resolved measurements of the THz electric field, enabling quantitative determination of the transient complex optical conductivity~\cite{Koch2023THzTDS}.

The ability of TRTS to characterize photoinduced states depends on how reliably their transient complex electrodynamic response can be determined across the relevant energy scales. Continuous phase-resolved coverage over a sufficiently broad spectral range enables the dynamics of responses at different energies to be compared under consistent excitation conditions and a common temporal reference, while helping distinguish microscopic processes that can appear similar within a limited spectral window~\cite{Zhang2024LBCO}. Adequate spectral resolution is needed to separate nearby excitations and track changes in their resonance frequencies and linewidths~\cite{zhangJosephsonNSR2023}. High sensitivity is required to resolve weak photoinduced responses, while long-term waveform stability supports reliable measurements during extended experiments~\cite{Huang2026THz2DCS,zhangPhotoinducedNonequilibrium2020}. Broadly tunable excitation further allows distinct nonequilibrium pathways to be selectively accessed~\cite{Disa2021Review,yang_terahertz_2023,Li2026StrongFieldTHz}.

Recent advances in millijoule-class Yb-based ultrafast lasers have enabled long-term operational stability at kilohertz repetition rates, providing sufficient pulse energy for both THz probing and flexible optical excitation~\cite{Donaldson2023Yb}. Multi-pass-cell post-compression further supplies the short pulses required for broadband THz generation and detection~\cite{Viotti2022MPC}. Achieving continuous phase-resolved spectral coverage within a single THz waveform scan, however, depends critically on the generation and detection schemes. Two-color gas plasmas provide exceptionally broad spectra, while their phase-sensitive generation can be susceptible to environmental fluctuations~\cite{Buldt2021,Mi2022}. Nonlinear crystals offer efficient THz generation and can support broad multi-THz spectra, although intrinsic phonon resonances remain imprinted on the spectral response~\cite{Jazbinsek2019OrganicTHz,Mansourzadeh2025,Cui2026BNA,Jeong2025BNA}. Spintronic THz emitters avoid phase-matching constraints, with the usable bandwidth largely determined by the detection scheme~\cite{Vogel2022}. Very thin electro-optic crystals can extend detection to higher frequencies, but internal reflections can introduce early echoes that shorten the usable temporal window and thereby limit spectral resolution~\cite{Kampfrath2007}. Combining measurements across separate spectral windows can extend the overall bandwidth, although differences in excitation conditions and pump–probe timing can complicate quantitative comparisons of transient responses.

Here, we combine multi-pass-cell pulse compression, spintronic THz generation, and air-biased coherent detection in a Yb-laser-based TRTS platform. A single THz time-domain waveform scan provides continuous phase-resolved spectral coverage to 25~THz, with a Fourier resolution of 0.1~THz and a time-domain signal-to-noise ratio of 72~dB, enabling ultrabroadband, high-sensitivity THz spectroscopy. Approximately 80\% of the laser pulse energy remains available for flexible excitation. Measurements on intrinsic GaAs demonstrate quantitative retrieval of equilibrium and transient complex optical properties, including under strongly depth-dependent photoexcitation where conventional approximate corrections fail. This platform substantially extends the range of nonequilibrium electrodynamic information that can be accessed directly and reliably within a single experiment.

\begin{figure}[htbp]
	\centering
	\includegraphics[width=0.96\linewidth]{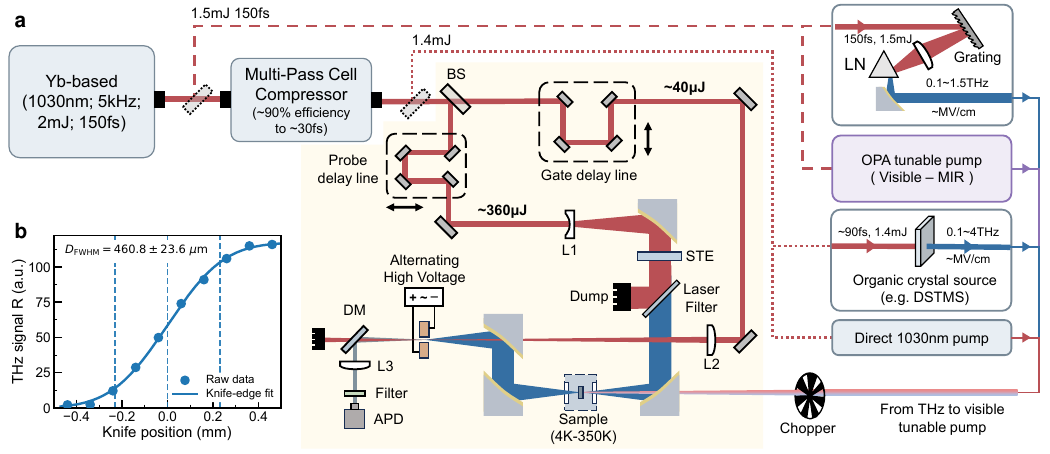}
	\caption{Yb-laser-based ultrabroadband TRTS platform. (a) Schematic of the pump--probe platform. The yellow shaded region denotes the ultrabroadband THz generation and detection modules. Most of the approximately 2-mJ pulse energy is available for excitation, including intense THz generation from LiNbO$_3$ (LN) or organic crystals and visible- to mid-infrared generation with an optical parametric amplifier (OPA). BS, beam splitter; STE, spintronic THz emitter; L1, concave lens; L2 and L3, convex lenses; DM, dichroic mirror; APD, avalanche photodiode. (b) Knife-edge characterization of the THz probe at the sample position, yielding a focal-spot FWHM of $460.8\pm23.6~\ensuremath{\mu}\mathrm{m}$.}
	\label{fig1}
\end{figure}

\section{Experimental Setup}
Figure~\ref{fig1}(a) shows the experimental layout. The system is built around a millijoule-class Yb-based femtosecond laser (YC-FL-10-2000-IR) operating at 5~kHz, with a center wavelength of 1030~nm, a pulse energy of approximately 2~mJ, and an initial pulse duration of $\sim$150~fs. The pulses are spectrally broadened in an Ar-filled multi-pass cell (MPC; YactoCompress) and subsequently compressed to $\sim$30~fs.

After compression, approximately 360~\ensuremath{\mu}J is used to drive a sapphire-supported W/CoFeB/Pt spintronic THz emitter (STE) for ultrabroadband THz generation~\cite{Seifert2016,Rouzegar2023}. A concave lens (L1) and a $90^{\circ}$ off-axis parabolic (OAP) mirror control the optical fluence on the emitter. Residual 1030-nm light transmitted through the STE is removed by laser-blocking elements. The THz probe is focused to a full width at half maximum (FWHM) of $\sim$460~\ensuremath{\mu}m at the sample position [Fig.~\ref{fig1}(b)], where a cryostat provides a temperature range of 4--350~K. A separate $\sim$40~\ensuremath{\mu}J branch serves as the optical gate for air-biased coherent detection (ABCD)~\cite{Dai2006,Zhou:26}. 

Approximately 80\% of the laser pulse energy remains available for pump excitation [Fig.~\ref{fig1}(a)]. This energy is sufficient to generate strong-field THz pulses, reaching peak electric fields above 1~MV/cm from LiNbO$_3$ and DSTMS crystals~\cite{Kong2025,Zheng2024}. It can also drive an optical parametric amplifier (OPA) for broadly tunable visible-to-mid-infrared excitation~\cite{Budriunas2022LongSeed}.

\section{Results}

\subsection{Ultrabroadband THz Probe Characterization}

\begin{figure}[htbp]
  \centering
  \includegraphics[width=0.96\linewidth]{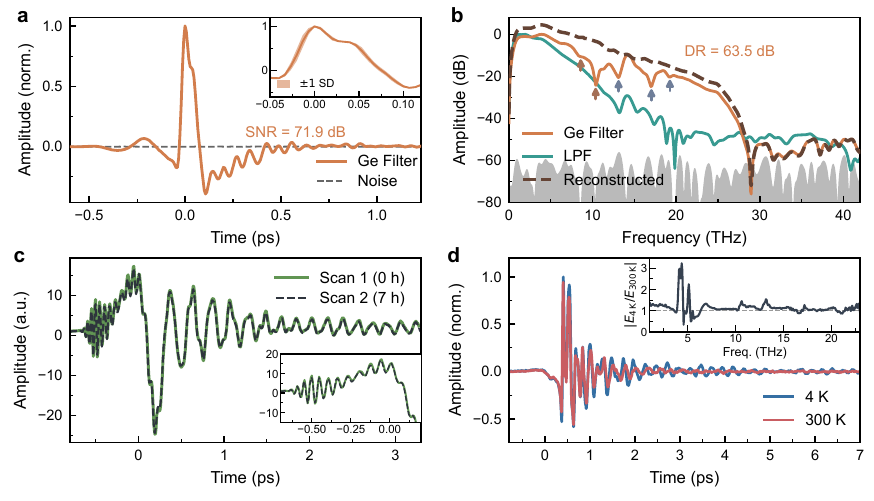}
  \caption{Characterization of THz probe. (a) Time-domain waveform measured using a 1-mm-thick Ge plate to block residual laser. Inset enlarges the waveform near 0~ps; shading indicates $\pm1$ standard deviation (SD) over five repeated measurements. (b) THz amplitude spectra using different filters. Arrows mark phonon-related features from Ge (brown) and sapphire (grayish blue); LPF denotes low-pass filter. The dashed curve is the spectrum with phonon responses de-embedded. (c) Time-domain waveforms measured on GaAs, separated by 7~h. Inset provides an enlarged view of the high-frequency oscillations at the leading edge of the waveforms. (d) Time-domain waveforms and corresponding spectral ratio of GeSe at 4~K and 300~K.}
  \label{fig2}
\end{figure}

Figure~\ref{fig2}(a) shows the measured THz waveform. A single scan reaches a time-domain signal-to-noise ratio (SNR) of 71.9~dB with a lock-in time constant of 200~ms, corresponding to a relative field noise of approximately $2.5\times10^{-4}$. The first internal-reflection echo appears at approximately 9.6~ps, in good agreement with the round-trip propagation time through the 430-\ensuremath{\mu}m sapphire substrate of the STE. The resulting echo-free temporal window corresponds to a Fourier spectral resolution of approximately 0.10~THz. The inset highlights the main pulse and its scan-to-scan reproducibility, with the shaded region representing $\pm1$ standard deviation over five repeated scans.

The corresponding amplitude spectrum [Fig.~\ref{fig2}(b)] reaches a peak dynamic range (DR) of 63.5~dB and remains nearly continuous to 28~THz. At 25~THz, the signal remains less than 30~dB below the spectral maximum, leaving substantial margin for phase-resolved spectroscopy. Spectral minima near 8.6, 10.4, 13, 17, and 19~THz arise mainly from phonon-related responses of the 1-mm-thick Ge blocking filter and the sapphire substrate of the STE~\cite{Shirley2007,Deinzer2004,Gueckstock2021}. A ceramic low-pass filter (LPF) gives a smoother low-frequency response but attenuates the high-frequency spectrum [Fig.~\ref{fig2}(b)]. To isolate the contributions of the Ge plate and sapphire substrate, their responses are de-embedded using literature optical constants~\cite{Loewenstein1973,Rouzegar2023,Mills1955,Burnett2020,Alishouse1966,Seifert2016,Seifert2022,Barker1963}. The resulting spectrum is shown by the dashed curve in Fig.~\ref{fig2}(b).

Long-term stability is evaluated by comparing room-temperature GaAs waveforms acquired 7~hours apart under identical conditions [Fig.~\ref{fig2}(c)]. The two traces agree within the experimental noise, including the $\sim$20 THz high-frequency oscillations at the leading edge of the THz waveform, as highlighted in the inset, showing no measurable amplitude or phase drift over this period. Temperature-dependent measurements are illustrated using GeSe, whose time- and frequency-domain responses at 300 and 4~K show clear changes in the phonon features [Fig.~\ref{fig2}(d)].

\subsection{Equilibrium optical constants of GaAs}

\begin{figure}[htbp]
  \centering
  \includegraphics[width=0.93\linewidth]{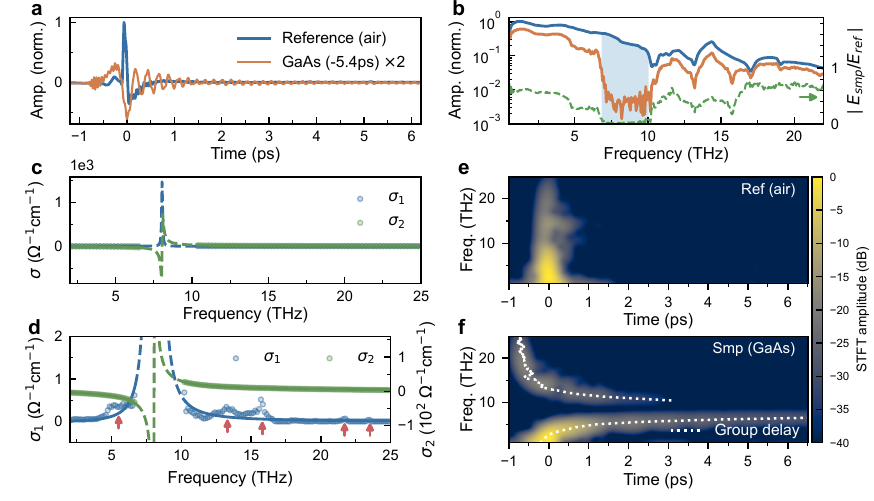}
  \caption{Equilibrium optical response of GaAs. (a) THz fields transmitted through air and GaAs; the GaAs trace is shifted by $-5.4$~ps and scaled by 2 for clarity. (b) Fourier-amplitude spectra and field-transmission ratio; the 7--10~THz minimum is associated with a TO phonon. (c) Retrieved complex optical conductivity. (d) Expanded weak-absorption region; solid curves are Drude--Lorentz fits outside the strong TO phonon absorption band, while dashed curves show the modeled response within the excluded region; arrows mark higher-order lattice features. (e,f) Short-time Fourier-transform maps for air and GaAs; the dashed curve in (f) is the group delay calculated from the retrieved refractive index.}
  \label{fig3}
\end{figure}

Figure~\ref{fig3}(a) shows the electric fields transmitted through air and GaAs, with the corresponding Fourier spectra in Fig.~\ref{fig3}(b). The complex refractive index $N(\omega)=n(\omega)+\mathrm{i}k(\omega)$ of GaAs is retrieved from the measured complex transmission ratio,
$H_{\exp}(\omega)=E_{\mathrm{smp}}(\omega)/E_{\mathrm{ref}}(\omega)$.
For each trial value of $N(\omega)$, we calculate the transmission of a 500-\ensuremath{\mu}m-thick GaAs slab from the Fresnel coefficients and propagation phase, and determine $N(\omega)$ by minimizing the complex difference from $H_{\exp}(\omega)$. The dielectric function and optical conductivity then follow directly~\cite{Duvillaret1996,Dorney2001,Jepsen2011}. The 7--10~THz near-zero-transmission region is excluded from the retrieval because of its high noise sensitivity. The localized spectral structures introduced by the Ge filter and sapphire substrate largely cancel in the sample-to-reference ratio and therefore do not compromise the broadband retrieval. Numerical details are given in Supporting Information Sec.~S1.

Figure~\ref{fig3}(c) shows the resulting complex optical conductivity, dominated by a transverse-optical (TO) phonon near 8~THz. The expanded scale in Fig.~\ref{fig3}(d) reveals much weaker higher-order lattice absorption. The step-like structure near 5--6~THz is consistent with second-order phonon difference-frequency absorption, while weaker features near 13.3, 15.8, 21.7, and 23.5~THz are associated with higher-frequency combination and multiphonon processes~\cite{Cochran1961,Komandin2024}. The observation of both the strong reststrahlen response and substantially weaker higher-order features across the measured spectrum demonstrates a nearly continuous spectroscopic coverage extending to 25~THz.

The pronounced frequency-dependent delay of the GaAs waveform provides an independent check of the retrieved refractive index. Figures~\ref{fig3}(e) and \ref{fig3}(f) compare short-time Fourier-transform (STFT) maps of the air-reference and GaAs waveforms. From the retrieved $N(\omega)$, we calculate the group refractive index $n_g(\omega)$ and the relative group delay
$
\Delta t_g(\omega)=\frac{d}{c}[n_g(\omega)-1],
$
where $d$ is the GaAs thickness. The calculated delay, shown by the white dashed curve in Fig.~\ref{fig3}(f), follows the measured temporal dispersion and provides an independent consistency check on the broadband optical constants.

\subsection{Transient photoconductivity of GaAs}

\begin{figure}[htbp]
  \centering
  \includegraphics[width=0.82\linewidth]{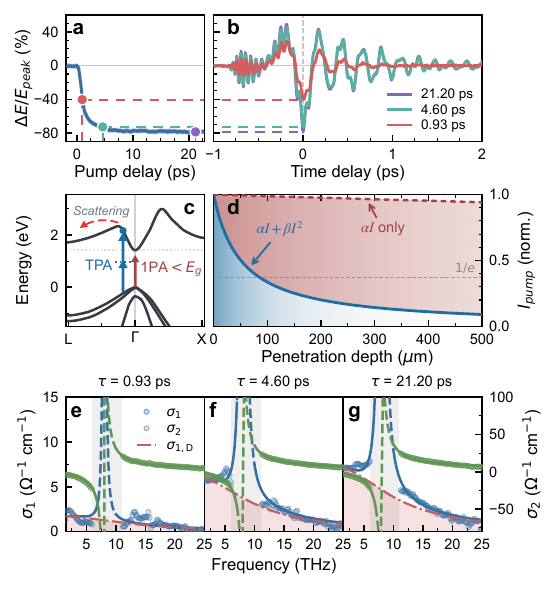}
  \caption{Ultrafast dynamics of GaAs after 1030-nm excitation.
(a) Normalized transient field change $\Delta E/E_{\mathrm{peak}}$ as a function of pump--probe delay.
(b) Transient THz waveforms $\Delta E/E_{\mathrm{peak}}$ at three representative delays.
(c) Schematic band structure of GaAs illustrating two-photon excitation (TPA), weak sub-bandgap one-photon absorption (1PA), and intervalley carrier scattering.
(d) Calculated depth dependence of the normalized pump intensity including both linear and two-photon absorption (blue solid line), compared with linear absorption alone (red dashed line). The gray dashed line marks the $1/e$ intensity level.
(e--g) Retrieved transient complex photoconductivity at $\tau=0.93$, 4.60, and 21.20~ps. The gray spectral interval is strongly suppressed by the TO phonon and excluded from the quantitative fit. Red shaded regions indicate the fitted Drude contribution to $\sigma_1(\omega)$.}
  \label{fig4}
\end{figure}

We then track the transient response of GaAs after excitation by a 1030-nm pulse at a fluence of $500~\mu\mathrm{J/cm^2}$. The pump--probe delay trace is acquired by synchronously scanning the probe and gate delay lines shown in Fig.~\ref{fig1}(a), thereby varying the pump--probe delay while maintaining a fixed sampling point on the THz waveform. In contrast, scanning only the probe delay line allows the transient THz waveform to be recorded at a selected pump--probe delay. Figure~\ref{fig4}(a) shows the pump-induced field change as a function of pump--probe delay, while full transient waveforms at representative delays are shown in Fig.~\ref{fig4}(b). The peak amplitudes extracted from the full waveforms agree with the corresponding points in the delay scan, confirming consistency between the two acquisition modes.

At room temperature, the 1030-nm pump photon energy lies below the direct GaAs band gap of $\sim1.43$~eV~\cite{Malone_2013}. Linear absorption is therefore weak~\cite{ValleyGaAs}, whereas two-photon absorption (TPA) provides an efficient excitation channel~\cite{Bechtel1976,Boggess1985,DeSalvo1993}. We describe the pump propagation by
\begin{equation}
\frac{\partial I(t,z)}{\partial z}
=-\alpha I(t,z)-\beta I^2(t,z),
\label{eq}
\end{equation}
where $\alpha$ and $\beta$ are the linear- and two-photon-absorption coefficients, respectively. As shown in Fig.~\ref{fig4}(d), the pump intensity decreases strongly with depth. Because the TPA contribution to the local carrier generation rate scales as $I^2$, the resulting photoexcited-carrier distribution is strongly weighted toward the incident surface, whereas the THz probe samples a much greater depth. This pronounced depth mismatch must therefore be included in the transient optical analysis.

Conventional penetration-depth corrections yield unphysical negative $\sigma_1(\omega)$ over part of the measured spectrum for the present data (Supporting Information Sec.~S2). The failure of these approximations arises from the simplified phase treatment used to obtain an analytic solution, which becomes inaccurate for the large photoinduced field change across such a broad spectral range~\cite{DAngelo2016}. We therefore retrieve the transient optical response using a multilayer transfer-matrix calculation that explicitly incorporates the depth-dependent excitation. The surface dielectric perturbation, $\Delta\widetilde{\epsilon}_{\mathrm{s}}(\omega)$, is first estimated through a coarse parametric fit and then refined variationally under Kramers--Kronig constraints~\cite{Karlsen2019, Kuzmenko2005,Ashoka2022}. Details of the excitation-profile calculation and inversion procedure are given in Supporting Information Sec.~S2.

Figures~\ref{fig4}(e)--\ref{fig4}(g) show the retrieved transient complex photoconductivity at representative pump delays. At $\tau=0.93$~ps, the fitted Drude response is relatively weak, but grows substantially by $\tau=4.60$~ps and changes only modestly thereafter up to 21.20~ps. This delayed buildup of low-frequency conductivity is consistent with two-photon excitation initially populating states well above the band edge, followed by intervalley scattering and energy relaxation before a stronger Drude response develops~\cite{Nuss1987,Shah1987,Kanasaki2014}. The Lorentz fits also reveal a broadening of the phonon resonance upon photoexcitation.

\section{Discussion}

The performance reached in this TRTS platform is deliberately matched to the characteristic energy and signal scales of low-energy spectroscopy in quantum materials. Continuous phase-resolved coverage to 25~THz ($\sim100$~meV) encompasses a broad range of single-particle and collective excitations~\cite{RevModPhys.93.041002}, while the 0.1 THz ($\sim0.4$ meV) Fourier resolution resolves the frequencies, linewidths, and transient shifts of many infrared-active resonances~\cite{Dong2023Recent}. The relative field noise of $\sim2.5\times10^{-4}$ further preserves sub-percent photoinduced changes across this broadband response~\cite{Demsar2020}. These capabilities make spectral-weight redistribution, scattering dynamics, and resonant responses at widely separated energies simultaneously accessible with full amplitude and phase information. Because the probe requires only a small fraction of the laser output, the same measurements can be combined with tunable strong-field excitations. Built on a commercially mature millijoule-class Yb laser system, the platform is broadly implementable across ultrafast laboratories, opening this spectroscopic capability to a much wider range of investigations.

The present configuration is designed as a broad-purpose operating point, with individual dimensions readily extendable for more specialized measurements. Suppressing or delaying the first internal-reflection echo, for example through substrate engineering, would extend the echo-free temporal window and further improve the spectral resolution for exceptionally narrow resonances~\cite{Kolejak2024}. Shorter drive and gate pulses could extend the usable probe bandwidth toward higher frequencies~\cite{Liu2026HighRepetitionRate}. For opaque or strongly absorbing materials that are inaccessible in the present transmission geometry, implementation in reflection geometry would extend the same broadband, phase-resolved capability to a substantially wider range of samples~\cite{Zhang2017Tunable}.

\section{Conclusions}

In conclusion, we establish a Yb-laser-based TRTS platform designed to provide a broad and quantitative view of nonequilibrium electrodynamics within a single experiment. It achieves continuous phase-resolved coverage to 25 THz ($\sim$100 meV), a Fourier spectral resolution of 0.1 THz ($\sim$0.4 meV), a single-scan time-domain SNR of 71.9 dB, and long-term amplitude and phase stability, while preserving most of the laser energy for broadly tunable excitation. Broadband measurements on intrinsic GaAs demonstrate accurate retrieval of both equilibrium and transient complex optical properties. By allowing spectral-weight redistribution, scattering dynamics, and resonant excitations across widely separated energy scales to be followed within the same phase-resolved measurement, the platform provides a more complete microscopic picture of photoinduced nonequilibrium states in quantum materials.

\section*{Acknowledgments}

We sincerely thank Prof. Nan-Lin Wang for fruitful discussions. This work was supported by the National Key Research and Development Program of China (Nos.~2024YFA1408703, 2021YFA1400204 and 2025YFE0202100) and the National Natural Science Foundation of China (Nos.~U24A2016, 12304184 and 12574164).

\section*{Supporting Information}

Details of the equilibrium optical-constant retrieval and the depth-resolved transient multilayer inversion are described in the accompanying Supporting Information (Sections S1 and S2).

\printbibliography

\end{document}